\documentclass[12pt]{article}
\usepackage[latin1]{inputenc}
\usepackage{amsmath}
\usepackage{amsfonts}
\usepackage{amssymb}
\usepackage{graphicx}
\usepackage{epstopdf}
\usepackage{color}
\usepackage{appendix}
\usepackage{csvsimple}
\usepackage{longtable}
\usepackage{chngpage}
\usepackage{subcaption}
\usepackage[export]{adjustbox}[2011/08/13]
\usepackage{mathrsfs}
\usepackage{a4wide}
\usepackage{psfrag}
\usepackage{multirow}
\usepackage{hhline}

\begin{document}

\title{\bf How Thermal Inflation can save 
Minimal Hybrid Inflation in Supergravity}

\author{Konstantinos Dimopoulos and Charlotte Owen\\
\\
{\small\em Consortium for Fundamental Physics, Physics Department,}\\
{\small\em Lancaster University, Lancaster LA1 4YB, UK}
\\
\vspace{-.4cm}
\\
{\small e-mails: {\tt k.dimopoulos1@lancaster.ac.uk}, \
{\tt c.owen@lancaster.ac.uk}}}
\vspace{1cm}

\maketitle

\begin{abstract}
Minimal hybrid inflation in supergravity has been ruled out by the 2015 Planck
observations because the spectral index of the produced curvature perturbation
falls outside observational bounds. To resurrect the model, a number of 
modifications have been put forward but many of them spoil the accidental 
cancellation that resolves the $\eta$-problem and require complicated 
K\"{a}hler constructions to counterbalance the lost cancellation. In contrast, 
in this paper the model is rendered viable by supplementing the scenario with a
brief period of thermal inflation, which follows the reheating of primordial 
inflation. The scalar field responsible for thermal inflation requires a large 
non-zero vacuum expectation value (VEV) and a flat potential. We investigate 
the VEV of such a flaton field and its subsequent effect on the inflationary 
observables. We find that, for large VEV, minimal hybrid inflation in 
supergravity produces a spectral index within the 1-sigma Planck bound and a 
tensor-to-scalar ratio which may be observable in the near future. The 
mechanism is applicable to other inflationary models.
\end{abstract}

\nopagebreak


\section{Introduction}

The precision of the latest observational data from the Planck satellite is
so high that it excludes several families of well motivated and thoroughly 
explored inflationary models \cite{Ade:2015lrj}. A prominent example is minimal
hybrid inflation. 

Hybrid inflation was introduced by Linde \cite{Linde:1991} to 
employ sub-Planckian field values whilst avoiding `unnaturally' tiny 
couplings. A supergravity (SUGRA) version of the model \cite{SUGRA-HI} has the 
neat feature that a minimal K\"{a}hler potential $K$, apart from rendering the 
fields canonically normalised, avoids excessive K\"{a}hler corrections due to 
an accidental cancellation.

Inflationary model-building in SUGRA suffers from the infamous 
$\eta$-problem, because the scalar potential is proportional to $e^{K/m_P^2}$ 
which results in \mbox{$\eta=K''+\cdots$}, where $\eta$ is the second slow-roll 
parameter and the primes denote derivatives with respect to the inflaton. For 
canonical fields \mbox{$K''=1$} so, in general, slow-roll is spoilt. 
However, for hybrid inflation with a minimal K\"{a}hler potential a term in the
ellipsis cancels $K''$ and allows \mbox{$\eta\ll 1$} during inflation.

It was rather disappointing that the Planck observations killed the minimal 
hybrid model. This is why many authors put forward modifications, which 
produce observables within the allowed ranges but at the expense of the above 
accidental cancellation. Thus, elaborate K\"{a}hler constructions were 
introduced to ensure slow-roll (e.g. see Ref.~\cite{Kahler} and references 
therein - but also see Ref.~\cite{apostolos}). 

Another proposal is Double Inflation \cite{double}, which retains the accidental
cancellation mentioned above, and utilises a second period of inflation, such 
that the number of e-folds of remaining primordial inflation when the 
cosmological scales exit the horizon is reduced. As a result, the inflationary 
observables are affected in a way that renders them compatible with 
observations. In Double Inflation, however, the inflationary model is modified
to enable the second stage of inflation (which follows directly after the first
stage), when the field follows a semi-shifted path in configuration space.
Thus, the model is no longer minimal.

In this paper we propose a mechanism which manages to render the model 
compatible with observations whilst retaining the neat feature of the 
accidental cancellation of minimal hybrid inflation in SUGRA. In contrast to
Ref.~\cite{double}, we do not modify the primordial inflation model.
We simply consider that, after reheating, there is a period of thermal 
inflation due to some flaton scalar field. The mechanism operates with any 
suitable flaton of mass $\sim\,$1~TeV. As with Double Inflation, our mechanism 
dilutes unwanted relics, possibly formed at the end of 
minimal hybrid inflation in SUGRA.

We use natural units, for which \mbox{$c=\hbar=1$} and \mbox{$m_P^{-2}=8\pi G$}, 
with \mbox{$m_P=2.43\times 10^{18}\,$GeV} being the reduced Planck mass.

\section{Minimal Hybrid Inflation in Supergravity}

Hybrid inflation in SUGRA is achieved with the superpotential:
\begin{equation}
W = \kappa S(\Phi\bar{\Phi}-M^{2})+\cdots
\label{eq:SUSYSuperPot}
\end{equation}
where the dots denote Planck-suppressed non-renormalisable operators, 
$\Phi,\bar\Phi$
are a pair of singlet lefthanded superfields and $S$ is a gauge singlet 
superfield which acts as the inflaton. The flatness of the inflationary 
trajectory is guaranteed by a U(1) R-symmetry on $S$. The parameters $\kappa$ 
and $M$ are made positive with field redefinitions, where 
\mbox{$M\sim 10^{16}\,$GeV} is the scale of a grand unified theory (GUT) and
\mbox{$\kappa\leq 1$} is a dimensionless coupling constant. The supersymmetric 
minimum is at \mbox{$\langle\Phi\rangle=\langle\bar\Phi\rangle=M$} and 
\mbox{$\langle S\rangle=0$}. We also consider a minimal
K\"{a}hler potential for the fields:
\begin{equation}
K=|\Phi|^2+|\bar\Phi|^2+|S|^2
\end{equation}
Then the F-term scalar potential is:
\begin{equation}
V_F=\kappa^2|M^2-\Phi\bar\Phi|^2+\kappa^2|S^2|(|\Phi|^2+|\bar\Phi|^2)+\cdots
\label{eq:HybridScalarPot}
\end{equation}
where the dots denote Planck-suppressed terms and we consider that the fields 
are sub-Planckian.

When \mbox{$|\langle\Phi\rangle|=|\langle\bar\Phi\rangle|$} the D-terms vanish.
Since the soft-breaking terms are negligible near the inflation scale (the GUT 
scale), the scalar potential is \mbox{$V=V_F+\Delta V$}, where $\Delta V$ is the
Coleman-Weinberg one-loop radiative correction:
\begin{equation}
\Delta V\simeq\frac{\kappa^4M^4}{16\pi^2}\ln\frac{\kappa^2|S|^2}{\Lambda^2}\,,
\label{DeltaV}
\end{equation}
where $\Lambda$ is some renormalisation scale. By suitable rotations in field 
space we write \mbox{$\sigma=\sqrt 2 S$} and \mbox{$\Phi=\bar\Phi=\varphi/2$}, 
where $\sigma$, $\varphi$ are canonically normalised real scalar fields 
\cite{Lazarides:2001}. 

From Eq.~\eqref{eq:HybridScalarPot}, the mass-squared for the waterfall 
field $\varphi$ is
\begin{equation}
m_\varphi^2\simeq\kappa^2(|S|^2-M^2)\,.
\end{equation}
Hence, when \mbox{$\sigma<\sigma_c\equiv\sqrt 2 M$}, the field becomes 
tachyonic. For \mbox{$\sigma>\sigma_c$} the potential is minimised for 
\mbox{$\Phi=\bar\Phi=0$} (i.e. \mbox{$\varphi=0$}) and 
inflation is driven by the false vacuum density $\kappa^2M^4$. Inflation ends 
abruptly at the waterfall when \mbox{$\sigma=\sigma_c$}. Then, during inflation,
the potential becomes 
\begin{equation}
V=\kappa^2M^4+\frac{\kappa^4M^4}{8\pi^2}
\ln\left(\frac{\kappa\sigma}{\sqrt 2\Lambda}\right)\,.
\label{HybridScalarPotential}
\end{equation}
where the second term is subdominant but provides a slope along the 
inflationary valley, necessary for slow-roll. One of the advantages of the 
model is that, by keeping the minimal K\"{a}hler potential, the accidental cancellation 
still avoids the $\eta$-problem, otherwise endemic in SUGRA inflation.

The slow-roll parameters are:
\begin{equation}
\epsilon=\frac{m_P^2}{2}\left(\frac{V'}{V}\right)^2= 
\frac{\kappa^4}{128\pi^4}\left(\frac{m_P}{\sigma}\right)^2\,,
\label{HybridEpsilon}
\end{equation}
\begin{equation}
\eta=m_P^2\frac{V''}{V}=-\frac{\kappa^2}{8\pi^2}
\left(\frac{m_P}{\sigma}\right)^2\,,
\label{HybridETa}
\end{equation}
where the prime denotes derivative with respect to the inflaton $\sigma$.
For 
the spectral index of the density 
perturbations we have
\begin{equation}
n_s - 1 =-\frac{\kappa^2}{4\pi^2}\left(1+\frac{3\kappa^2}{16\pi^2}\right)
\left(\frac{m_P}{\sigma}\right)^2\,,
\end{equation}
while the tensor to scalar ratio is given by the consistency condition as 
\mbox{$r=16\epsilon$}. Expressing the above in terms of the remaining inflation
e-folds $N$, using
\begin{equation}
\Big(\frac{\sigma}{m_P}\Big)^2 = 2\Big(\frac{M}{m_P}\Big)^2+
\frac{\kappa^2}{4\pi^2}N\,,
\label{eq:sigma}
\end{equation}
where \mbox{$\sigma_c^2\equiv 2M^2$}, we find:
\begin{equation}
r=16\epsilon
\simeq\frac{\kappa^2}{2\pi^2}\frac{1}{N}
\label{Hybridr}
\end{equation}
and
\begin{equation}
n_s-1=2\eta-6\epsilon\simeq-\frac{1}{N}\,.
\label{Hybridns}
\end{equation}

Examining Eqs. \eqref{Hybridr} and \eqref{Hybridns} with 
\mbox{$\kappa=0.1$} and \mbox{$N=60\;(50)$} gives us 
\mbox{$r=8\times 10^{-6}$} \mbox{$(1\times 10^{-5})$} and 
\mbox{$n_s=0.983\;(0.980)$}. Whilst $r$ is well beneath the Planck upper bound, 
but as yet unobservable, $n_s$ is clearly above the upper 2-$\sigma$ bound of 
the Planck observations \cite{Ade:2015lrj}. Therefore, the model appears to be
excluded. To account for this problem, many authors suggested modifications of
the theoretical setup. However, these lost the accidental cancellation that 
resolved the $\eta$-problem. Consequently, this required the further 
introduction of complicated versions of the K\"{a}hler potential to keep 
K\"{a}hler corrections under control. In this paper, we propose a simpler 
tactic. By examining Eqs.~\eqref{Hybridr} and \eqref{Hybridns} we notice that a 
lower value of $N$ can bring $n_s$ within the Planck bounds. Furthermore,
it may also increase $r$ to the point of observability.

\section{Thermal Inflation} 

\subsection{The model}

We will assume the existence of a so-called flaton field, which can
realise thermal inflation. This is a brief period of inflation lasting up to 
\mbox{$N_T\simeq 17$} e-folds, occurring shortly after reheating from 
primordial inflation.\footnote{%
It may occur even earlier, but we will not consider this possibility here.}
A bout of thermal inflation occurs from the moment when the scalar potential
dominates the thermal bath density, with the field being held on top of a
false vacuum, until the moment when a phase transition releases the field from
the false vacuum and sends it towards its true vacuum expectation value (VEV). 
Therefore, thermal inflation requires a scalar field with a large VEV and a 
flat potential; such a field is called a flaton~\cite{Lyth:1995}. Typically, a 
flaton field corresponds to a supersymmetric flat direction, lifted by a soft 
mass $\sim\,$1~TeV.

Now, the number of e-folds before the end of 
primordial inflation that correspond to the exit of the cosmological scales, 
$N$, can be reduced by $N_T$. By investigating the effect of the VEV of the 
thermal flaton on $N_T$ we can determine $N$ anew and hence  $r(N)$ and 
$n_s(N)$ for inflationary models. We find that this may render minimal hybrid 
inflation in SUGRA viable.

The zero temperature flaton potential and its VEV are:
\begin{equation}
V=V_0-\frac12m^2\phi^2+
\frac{\lambda}{[2(n+2)]!}
\frac{\phi^{2(n+2)}}{m_P^{2n}}\,,
\label{eq:Flaton}
\end{equation}
\begin{equation}
\langle\phi\rangle = \Big[\frac{\lambda}{(2n+3)!}\Big]^{-\frac{1}{2(n+1)}}
(m_P^nm)^{\frac{1}{n+1}}\,,
\label{eq:VEV}
\end{equation}
where $n>0$ is an integer. 
For a given $m$, a large VEV is attained with large $n$. When
\mbox{$n\rightarrow\infty$} we have \mbox{$\langle\phi\rangle\rightarrow m_P$}.
This case may correspond to a string modulus, whose otherwise flat potential 
may vary over $m_P$ distances in field space~\cite{Lyth:1995}. 

From the above, it is easy to show that the flaton mass-squared is
\begin{equation}
V''(\langle\phi\rangle)=2(n+1)m^2
\label{mass2}
\end{equation}
Also, because $V(\langle\phi\rangle)=0$ we find:
\begin{equation}
V_0 =\frac12\Big(\frac{n+1}{n+2}\Big)m^2\langle\phi\rangle^2\,.
\label{eq:V0}
\end{equation}
Thus, the energy scale of thermal inflation is, 
\mbox{$V_{0}^{1/4}\sim\sqrt{m\langle\phi\rangle}$}. 

However, the flaton field interacts strongly with the thermal bath, which 
introduces an additional, sizeable, term in Eq.~\eqref{eq:Flaton} of the form
\mbox{$\frac12 g^2T^2\phi^2$}, where $g$ is the coupling to the thermal bath 
and $T$ is the temperature~\cite{Lyth:1995}. Thus, the effective mass-squared 
is now
\begin{equation}
m_{\rm eff}^2=g^2T^2-m^2
\label{meff}
\end{equation}
For high temperatures, $m_{\rm eff}^2$ is positive and the flaton is driven to 
zero, where the potential in Eq.~\eqref{eq:Flaton} is \mbox{$V(0)=V_0$}.
Thermal inflation begins when $V_0$ overwhelms the density of the thermal bath,
\mbox{$\rho_T=\frac{\pi^2}{30}g_*T^4$}, where $g_*$ is the effective 
relativistic degrees of freedom. Thus, the temperature at the onset of thermal 
inflation is
\begin{equation}
T_1=\Big(\frac{30}{\pi^2g_{*}}\Big)^{\frac{1}{4}}V_{0}^{1/4}\sim V_{0}^{1/4}\,.
\label{eq:thermaldensity}
\end{equation}
Thermal inflation continues as long as the flaton field is kept on top of the 
false vacuum by the thermal interaction term. However, the thermal bath is 
exponentially depleted, so the temperature drops enough that $m_{\rm eff}^2$
becomes tachyonic and the flaton is released towards its VEV. This event 
terminates thermal inflation, and corresponds to the temperature
\begin{equation}
T_2=m/g\,.
\end{equation}
The e-folds of thermal inflation are
\begin{equation}
N_T
=\ln\Big(\frac{T_1}{T_2}\Big)
\simeq\ln\frac{gV_{0}^{1/4}}{m}\simeq
\frac12\ln\left(\frac{\langle\phi\rangle}{m}\right)\,,
\label{eq:thermal1}
\end{equation}
where we considered $g\sim 1$. From the above we see that the flaton VEV 
determines $N_T$.

\subsection{Some particulars of Thermal Inflation}

Before continuing, we briefly discuss some particular issues regarding 
thermal inflation. First, we consider whether the flaton potential is 
given by Eq.~\eqref{eq:Flaton} during inflation as well as in the vacuum.
We also comment on the expectation value of the flaton field during and after 
inflation.

If the flaton were light (with sub-Hubble mass) during inflation then
it would undergo particle production and develop a non-zero condensate. After 
the end of inflation and before reheating, however, while the inflaton field is
oscillating, there is a thermal bath with temperature
\mbox{$T\sim(m_P^2\Gamma H)^{1/4}$} \cite{KT}, which is bigger that the 
reheating temperature $\sim\sqrt{m_P\Gamma}$ ($\Gamma$ being the inflaton 
perturbative decay rate) that reduces in time as $T\propto t^{-1/4}$, while 
$H\propto 1/t$. The thermal mass of the flaton is $\sim T$ (taking 
\mbox{$g\sim 1$}), which means that the field becomes heavy (with super-Hubble 
mass) and rolls towards zero when the Hubble parameter is 
\mbox{$H_x\sim(m_P^2\Gamma)^{1/3}>\Gamma$}, i.e. well before reheating, as it is
straightforward to~show. 

In SUGRA theories, however, a scalar field receives a 
mass of order $H$ \cite{DRT}, which is the source of the $\eta$-problem. This 
means that it is not light during inflation, but rolls to zero well before 
inflation ends. The Hubble-scale mass becomes negligible before reheating (when
$H(t)=H_x$) and is irrelevant during thermal inflation.

Another issue is the following.
One might worry that the coupling of the flaton to the thermal bath, which also
couples to the waterfall field, may affect the potential in inflation. Note, 
however, that the waterfall field is zero during inflation, as most probably is
the flaton as well (due to SUGRA corrections), while the particles of the 
thermal bath are absent because the latter is inflated away. Thus, we do not
expect the potential in inflation to be affected by any non-zero contributions 
due to the flaton field.

Finally, the phase transition which ends thermal inflation can produce unwanted 
dangerous relics such as domain walls. One way to circumvent this problem is to 
gauge the thermal flaton such that the end of thermal inflation produces only 
harmless local cosmic strings, whose energy scale 
$V_0^{1/4}$ is too low to have any cosmological significance. The U(1) 
Goldstone boson is eaten by the gauge field, which promptly decays assuming a 
large mass due to the VEV of the flaton.

\subsection{\boldmath The impact on $N$}

To find the impact 
on $N$ we need to consider when reheating occurred. The remaining number of 
e-folds of primordial inflation when the cosmological scales exit the horizon is
\begin{equation}
N=62.8+ \ln\Big(\frac{k}{a_0H_0}\Big)+
\frac13\ln\Big(\frac{g_*}{106.75}\Big)+
\ln\Big(\frac{V_{\rm end}^{1/4}}{10^{16}\,{\rm GeV}}\Big)-\Delta N-N_T\,,
\label{eq:N*}
\end{equation}
where $k=0.05\,$Mpc$^{-1}$ is the pivot scale and 
$(a_0H_0)^{-1}$ is the comoving Hubble radius today. 
In the above, $V_{\rm end}^{1/4}$ is 
the energy scale at the end of inflation (typically the GUT scale) 
and 
\begin{equation}
\Delta N=\frac13\ln\left(\frac{V_{\rm end}^{1/4}}{T_{\rm reh}}\right)\,,
\label{eq:DeltaN1}
\end{equation}
which is introduced by assuming that, between the end of inflation and 
reheating, the Universe is dominated by the inflaton condensate, that coherently
oscillates in a quadratic potential, near its global minimum. 
Inefficient reheating can lead to low values for $T_{\rm reh}$. 

To minimise $N$, we assume the lowest possible reheating temperature.
With a subsequent period of thermal inflation, the reheating from primordial 
inflation must have completed before the thermal inflation initiates. Hence, 
\mbox{$T_{\rm reh}^4\geq V_0$}. 
Therefore, the minimum $N$ is achieved when
\begin{equation}
\Delta N\simeq\frac13
\ln\left(\frac{V_{\rm end}^{1/4}}{\sqrt{m\langle\phi\rangle}}\right)\,,
\label{eq:DeltaN2}
\end{equation}

In view of the above, $N$ can be minimized by the combined effect of low 
reheating and subsequent thermal inflation by the number of e-folds given by
\begin{equation}
\Delta N+N_T\simeq\frac13
\ln\left(\frac{V_{\rm end}^{1/4}\langle\phi\rangle }{m^2}\right)\simeq
\frac13
\ln\left[\frac{V_{\rm end}^{1/4}}{m}\left(\frac{m_P}{m}\right)^{\frac{n}{n+1}}\right]
\label{dN}
\end{equation}
where we used that \mbox{$\langle\phi\rangle\sim(m_P^nm)^{1/(n+1)}$} in accordance
to Eq.~\eqref{eq:VEV} (we assumed $\lambda\sim 1$).

\section{\boldmath Impact of Varying 
$\langle\phi\rangle$ on Hybrid Inflation}

Eq.~\eqref{dN} suggests that to minimise $N$ we need to reduce both $m$ and $n$
as much as possible. We take the value $m=1$TeV for the flaton tachyonic 
mass. This results in a a mass $\geq 2$~TeV as suggested by Eq.~\eqref{mass2}
for $n\geq 1$, which is the lowest allowed by LHC. 

\begin{figure}[h]
\centering
\includegraphics[width=\columnwidth]{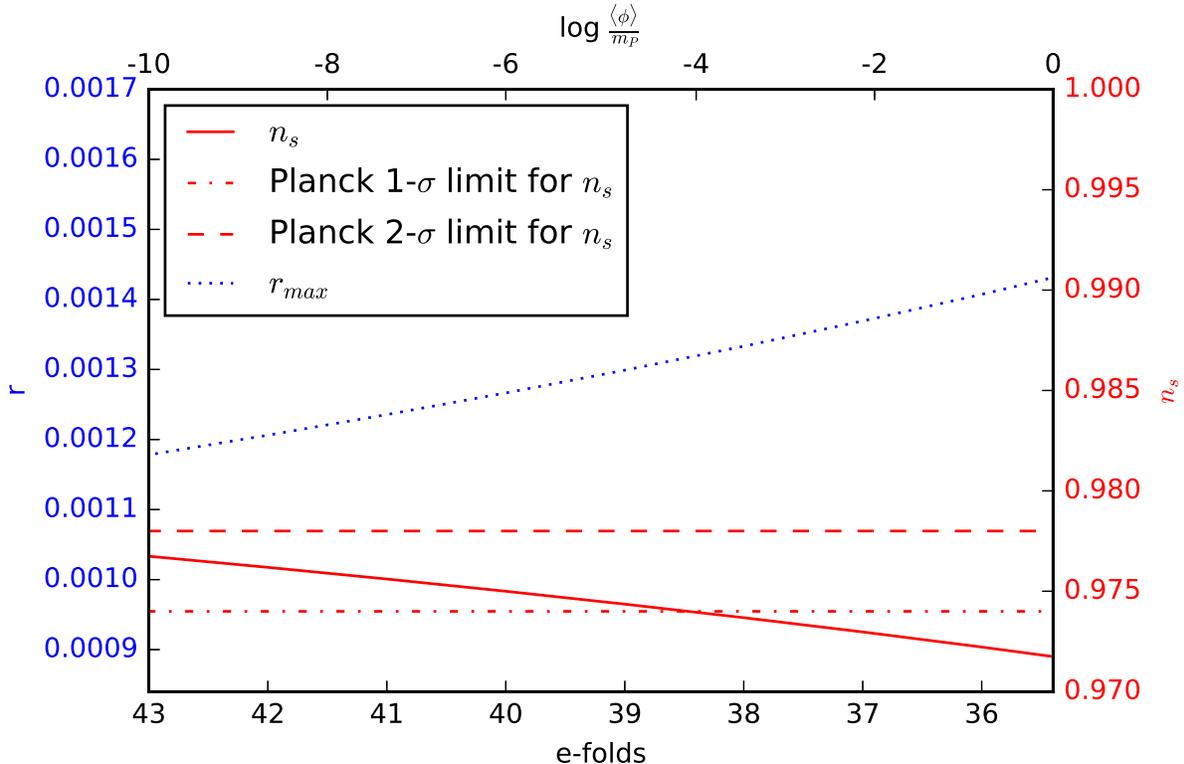}
\caption{$n_s$ and $r_{\rm max}$ in hybrid inflation as a function of 
$\langle\phi\rangle$ and therefore $N$. The solid line depicts the values of 
$n_s$. The horizontal dashed lines correspond to the bounds from the 2015 
Planck data on the spectral index (taken from Ref.~\cite{Ade:2015lrj}); 
the short-dashed line depicts the upper bound at 68\%~C.L. 
and the long-dashed line depicts the upper bound at 95\%~C.L..
The dotted line depicts the maximum allowed value for $r$, which 
corresponds to $\kappa=1$. Note that $n$ and $\lambda$ in Eq.~\eqref{eq:VEV} can
both vary and hence the value of the VEV falls on a continuum. }
\label{fig:FinalGraph}
\end{figure}

Table~\ref{table:FinalNvaluesfromorderofmagVEV} shows the results of varying 
$\langle\phi\rangle$, thereby determining the range of $N$ which can be used to 
calculate $r$ and $n_s$ for given inflationary models. 
\begin{table}[h]
\begin{center}
\begin{tabular}{|c|c|c|c|}
         \hline $\langle\phi\rangle$ (GeV) & $N_{T}$ & $N+N_T$ & $N$\\
         \hline $m_P$    & 17.7    & 53.1 & 35.4 \\
         \hline $10^{16}$    & 15.4    & 52.3 & 36.9 \\
         \hline $10^{14}$    & 13.1    & 51.6 & 38.4 \\
         \hline $10^{12}$    & 10.8    & 50.8 & 40.0 \\
         \hline $10^{10}$    & 8.5     & 50.0 & 41.5 \\
         \hline
\end{tabular} 
\end{center}
\caption{$N_T$, $N+N_T$ and $N$ with respect to~$\langle\phi\rangle$.}
\label{table:FinalNvaluesfromorderofmagVEV}
\end{table}
Focusing on SUGRA hybrid inflation and
using Eqs. \eqref{Hybridr} and \eqref{Hybridns} we find the results shown in 
Table~\ref{table:Finalnsandr} and Fig.~\ref{fig:FinalGraph}.

\begin{table}[h]
\begin{center}
\begin{tabular}{|c|c|c|c|}\hline
         $\langle\phi\rangle$ & $N$ & $n_s$ & $r_{\rm max}$
         \\ \hline
         $m_P$ & 35.4 & 0.972 & 1.4$\times 10^{-3}$
         \\ \hline
         $10^{10}\,$GeV & 41.5 & 0.976 & 1.2$\times 10^{-3}$
         \\ \hline
\end{tabular}
\end{center}
\caption{$n_s$ and $r_{\rm max}$ in hybrid 
inflation, with minimum $N$ and $\kappa=1$ (for maximum $r$).}
\label{table:Finalnsandr}
\end{table}


Fig. \ref{fig:FinalGraph} clearly shows our range of values for $n_s$ inside 
the parameter space of the Planck results, at the 2-$\sigma$ level. 
For large values of the VEV; i.e. more than $10^{13}\,$GeV or so, $n_s$ enters
inside the 1-$\sigma$ region. This corresponds to 
$n\geq 2$ in Eq.~\eqref{eq:VEV}, which, in view of Eq.~\eqref{mass2}, suggests
that the flaton mass is $\geq 2.4\,$TeV, which is safely above the latest LHC 
bounds.\footnote{%
Even though the quartic, self-interaction  term is absent for flaton fields by
definition \cite{Lyth:1995} (note also that, in supersymmetric theories, loop 
corrections logarithmically affect the mass term only \cite{LythRiotto}), we 
seem to require that the $\phi^6$ term is also suppressed 
for $n_s$ to lie inside the 1-$\sigma$ region of the Planck data, 
which amounts to some tuning.}

\section{Conclusions}

In conclusion, we have demonstrated that late reheating followed by a 
subsequent period of thermal inflation can enable the minimal hybrid inflation
in supergravity model to successfully produce cosmological perturbations with 
spectral index allowed by the Planck satellite observations. We have achieved 
this without affecting or modifying in the least the inflationary model, 
retaining thereby the accidental calcellation that resolves the $\eta$-problem
of inflation in supergravity with minimal K\"{a}hler potential. Furthermore, 
we also found that the tensor to scalar ratio can be significantly increased 
such that it may become observable in the near future. It is important to stress
that the above mechanism may also have profound implications for other 
inflationary models, see for example Ref.~\cite{plateau}. 

\section*{Acknowledgements}

KD would like to thank 
G.~Lazarides for discussions. 
CO is supported by the FST of Lancaster University. 
KD is supported (in part) by the Lancaster-Manchester-Sheffield 
Consortium for Fundamental Physics under STFC grant: 
ST/L000520/1.




\begin{thebibliography}{10}

\bibitem{Ade:2015lrj}
  P.~A.~R.~Ade {\it et al.} [Planck Collaboration],
  arXiv:1502.02114 [astro-ph.CO].

\bibitem{Linde:1991}
  A.~D.~Linde,
  Phys.\ Lett. \ B {\bf 259} {1991} 38;
  Phys.\ Rev.\ D {\bf 49} (1994) 748.

\bibitem{SUGRA-HI}
A.~D.~Linde and A.~Riotto,
  Phys.\ Rev.\ D {\bf 56} (1997) 1841;
G.~R.~Dvali, G.~Lazarides and Q.~Shafi,
  Phys.\ Lett.\ B {\bf 424} (1998) 259.

\bibitem{Kahler}
C.~Pallis and Q.~Shafi,
  Phys.\ Lett.\ B {\bf 736} (2014) 261;
  R.~Armillis, G.~Lazarides and C.~Pallis,
Phys.\ Rev.\ D {\bf 89} (2014) no.6,  065032;
  M.~U.~Rehman, Q.~Shafi and J.~R.~Wickman,
Phys.\ Rev.\ D {\bf 83} (2011) 067304.

\bibitem{apostolos}
B.~Garbrecht, C.~Pallis and A.~Pilaftsis,
JHEP {\bf 0612} (2006) 038.

\bibitem{double}
  G.~Lazarides and C.~Panagiotakopoulos,
Phys.\ Rev.\ D {\bf 92} (2015) no.12,  123502;
G.~Lazarides and A.~Vamvasakis,
Phys.\ Rev.\ D {\bf 76} (2007) 123514.

\bibitem{Lazarides:2001}
    G.~Lazarides,
    Lect.\ Notes Phys.\  {\bf 592} (2002) 351.
  
  \bibitem{Lyth:1995}
  D.~Lyth, E.~Stewart,
   Phys.\ Rev.\ D {\bf 53} (1995) 1784 

\bibitem{KT}
E.~W.~Kolb and M.~S.~Turner,
  Front.\ Phys.\  {\bf 69} (1990) 1.

\bibitem{DRT}
M.~Dine, L.~Randall and S.~D.~Thomas,
  Nucl.\ Phys.\ B {\bf 458} (1996) 291;
  Phys.\ Rev.\ Lett.\  {\bf 75} (1995) 398.

\bibitem{LythRiotto}
D.~H.~Lyth and A.~Riotto,
  Phys.\ Rept.\  {\bf 314} (1999) 1.

\bibitem{plateau}
K.~Dimopoulos and C.~Owen,
  arXiv:1607.02469 [hep-ph].

\end{thebibliography}
\end{document}